\documentstyle[aps,prl,epsfig,floats]{revtex}

\draft

\begin{document}
\twocolumn[\hsize\textwidth\columnwidth\hsize\csname
@twocolumnfalse\endcsname
\title{Comment on Neutrino Radiative Decay Limits from 
the Infrared Background}

\author{Georg G.~Raffelt}
\address{Max-Planck-Institut f\"ur Physik 
(Werner-Heisenberg-Institut),
F\"ohringer Ring 6, 80805 M\"unchen, Germany\\ \hbox{\ }}

\date{21 July 1998, submitted to Physical Review Letters}

\maketitle

\vskip2.0pc]

%%%%%%%%%%%%%%%%%%%%%%%%%%%%%%%%%%%%%%%%%%%%%%%%%%%%%%%%%%%%%%%%%%%%%%
%% Section I %%%%%%%%%%%%%%%%%%%%%%%%%%%%%%%%%%%%%%%%%%%%%%%%%%%%%%%%%
%%%%%%%%%%%%%%%%%%%%%%%%%%%%%%%%%%%%%%%%%%%%%%%%%%%%%%%%%%%%%%%%%%%%%%

Recent observations of TeV $\gamma$ rays from the active galaxies
Markarian 421 and 501 have provided new limits on the cosmic
background of infrared photons which are an efficient opacity source
due to the pair process $\gamma_{\rm TeV}\gamma_{\rm infrared} \to
e^+e^-$.  In a recent Letter {\em New Limits to the Infrared
Background: Bounds on Radiative Neutrino Decay and on Contributions of
Very Massive Objects to the Dark Matter Problem\/}~\cite{Biller} these
limits were interpreted, inter alia, as limits on the radiative decay
of cosmic background neutrinos with sub-eV masses.  However, it was
overlooked that in this mass range much more restrictive limits on
neutrino radiative decay channels already exist, preventing neutrino
decays from contributing significantly to the infrared background.

\begin{figure}[ht]
\hbox to\hsize{\hss\epsfxsize=8cm\epsfbox{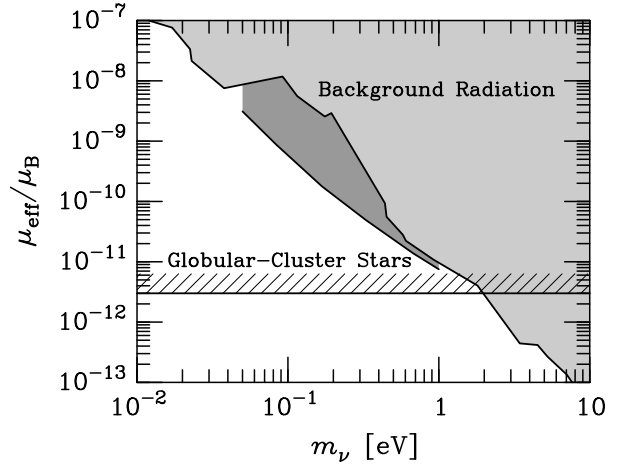}\hss}
\smallskip
\caption{Exclusion range for neutrino transition moments. The
light-shaded region was previously excluded from the contribution of
radiatively decaying neutrinos to the cosmic photon background
radiations~\protect\cite{Ressell}, the dark-shaded region is the new
exclusion range from the infrared background, case
10~TeV~\protect\cite{Biller}.  Values above the hatched bar are
excluded from the plasmon decay process in globular-cluster
stars~\protect\cite{Raffelt,Catelan,RaffeltBook}.
\label{fig:limits}}
\end{figure}

Neutrinos can have a variety of different electromagnetic form
factors. A radiative decay process $\nu_2\to\nu_1\gamma$ (masses $m_2$
and $m_1$, respectively) is uniquely characterized by the magnetic and
electric transition moments $\mu_{21}$ and $\epsilon_{21}$ according
to
\begin{eqnarray}\label{eq:decay}
\Gamma_{\nu_2\to\nu_1\gamma}&=&
\frac{|\mu_{21}|^2+|\epsilon_{21}|^2}{8\pi}\,
\left(\frac{m_2^2-m_1^2}{m_2}\right)^3\nonumber\\
&=&5.308~{\rm s}^{-1}\,
\left(\frac{\mu_{\rm eff}}{\mu_{\rm B}}\right)^2
\delta_m^3\,m_{\rm eV}^3,
\end{eqnarray}
where $\mu_{\rm eff}^2=|\mu_{21}|^2+|\epsilon_{21}|^2$, $\mu_{\rm
B}=e/2m_e$ is the Bohr magneton, $m_{\rm eV}=m_2/{\rm eV}$, and
$\delta_m=(m_2^2-m_1^2)/m_2^2$. This latter quantity is essentially
unity unless the neutrino masses are nearly degenerate.  The neutrino
radiative lifetime limits based on the cosmic photon
backgrounds~\cite{Biller,Ressell} thus provide exclusion regions in
the $\mu_{\rm eff}$-$m_\nu$-plane (Fig.~\ref{fig:limits}), assuming a
mass hierarchy $m_\nu=m_2\gg m_1$, i.e.~taking
$\delta_m\approx1$.

Neutrino dipole and transition moments are also constrained to avoid
excessive stellar energy-losses by the plasmon decay process
$\gamma\to\nu\bar\nu$, an idea going back to Ref.~\cite{Bernstein},
with more recent refinements in
Refs.~\cite{Raffelt,Catelan,RaffeltBook}. For neutrino masses below a
few keV the limit is $\mu_{\rm eff}\alt 3\times10^{-12}\,\mu_{\rm B}$,
shown as a hatched bar across Fig.~\ref{fig:limits}.  (For Dirac
neutrinos the limit would be slightly more restrictive because the
number of final states in $\gamma\to\nu\bar\nu$ doubles.)  The
$m_\nu^3$ phase-space factor in Eq.~(\ref{eq:decay}) is so punishing
that the direct radiative decay limits quickly become irrelevant for
$m_\nu\alt 2~{\rm eV}$.
The recent indications for neutrino masses from the solar and
atmospheric neutrino anomalies, the LSND experiment, and the
cosmological hypothesis of hot plus cold dark matter suggest that
neutrino masses are indeed below 1--2~eV. 

Therefore, unless something is unexpectedly wrong with the
globular-cluster limit, radiative neutrino decays cannot produce
infrared background photons at the level of current detection limits.
Interpreting measurements of photon backgrounds as constraints on
radiative neutrino decays is still important, of course, as it is an
independent method with other experimental and theoretical
uncertainties than the stellar-evolution arguments.

I thank Dr.~S.~Biller (for the authors of Ref.~\cite{Biller}) for a
collegial correspondence where they express agreement with my
analysis.  Partial support by the Deutsche
For\-schungs\-ge\-mein\-schaft under grant No.\ SFB~375 is
acknowledged.

%%%%%%%%%%%%%%%%%%%%%%%%%%%%%%%%%%%%%%%%%%%%%%%%%%%%%%%%%%%%%%%%%%%%%%
%% References %%%%%%%%%%%%%%%%%%%%%%%%%%%%%%%%%%%%%%%%%%%%%%%%%%%%%%%%
%%%%%%%%%%%%%%%%%%%%%%%%%%%%%%%%%%%%%%%%%%%%%%%%%%%%%%%%%%%%%%%%%%%%%%


\vskip-0.2cm

\begin{references}

\vbox to-1.6 cm{\vskip0pt minus 20cm}

\bibitem{Biller}
   S.D.~Biller et al., Phys. Rev. Lett. {\bf 80}, 2992 (1998).

\bibitem{Ressell}
  M.T.~Ressell and M.S.~Turner, Comm. Astrophys. {\bf 14}, 323 (1990).

\bibitem{Bernstein}
  J.~Bernstein, M.A.~Ruderman and G.~Feinberg, 
  Phys. Rev. {\bf 132}, 1227 (1963). 

\bibitem{Raffelt}
  G.G.~Raffelt, Astrophys. J. {\bf 365}, 559 (1990).
  Phys. Rev. Lett. {\bf 64}, 2856 (1990). 

\bibitem{Catelan}
  M.~Catelan, J.A.~de Freitas Pacheco and J.E.~Horvath,
  Astrophys. J. {\bf 461}, 231 (1996).

\bibitem{RaffeltBook}
  G.~Raffelt, {\it Stars as Laboratories for Fundamental Phys\-ics},
  (Chicago University Press, Chicago, 1996). 

\end{references}
\end{document}